# Spatial cancer systems biology resolves heterotypic interactions and identifies disruption of spatial hierarchy as a pathological driver event

*Fabian V. Filipp*[*]

[*] *Cancer Systems Biology, Institute of Diabetes and Cancer, Helmholtz Zentrum München, Ingolstädter Landstraße 1, D-85764 München, Germany*

[*] *School of Life Sciences Weihenstephan, Technical University München, Maximus-von-Imhof-Forum 3, D-85354 Freising, Germany*

[*] *Institute for Advanced Study, Technical University München, Maximus-von-Imhof-Forum 3, D-85354 Freising, Germany*

[*] *Phone: +49-89-3187-43455*

[*] *Email: fabian.filipp@helmholtz-muenchen.de*

[*] *ORCID: orcid.org/0000-0001-9889-5727*

[*] *Twitter: twitter.com@cancersystembio*

**Abstract**

Spatially annotated single-cell datasets provide unprecedented opportunities to dissect cell-cell communication in development and disease. Heterotypic signaling includes interactions between different cell types and is well established in tissue development and spatial organization. Epithelial organization requires several different programs that are tightly regulated. Planar cell polarity is the organization of epithelial cells along the planar axis orthogonal to the apical-basal axis. In this study, we investigate planar cell polarity factors and explore the implications of developmental regulators as malignant drivers. Utilizing cancer systems biology analysis, we derive gene expression network for WNT-ligands (WNT) and their cognate frizzled (FZD) receptors in skin cutaneous melanoma. The profiles supported by unsupervised clustering of multiple-sequence alignments identify ligand-independent signaling and implications for metastatic progression based on the underpinning developmental spatial program. Omics studies and spatial biology connect developmental programs with oncological events and explain key spatial features of metastatic aggressiveness. Dysregulation of prominent planar cell polarity factors such specific representative of the WNT and FZD families in malignant melanoma recapitulates the development program of normal melanocytes but in an uncontrolled and disorganized fashion.

**Introduction**

The emerging fields of single-cell RNA-sequencing (scRNA-Seq) and spatial transcriptomics (spRNA-Seq) combine the benefits of traditional histopathology with single-cell gene expression profiling. The ability to connect the spatial organization of molecules in cells and tissues with their gene expression state enables the mapping of developmental stages as well as resolving specific disease pathologies. spRNA-Seq has the ability to decode molecular proximities from sequencing information and construct images of gene transcripts at sub-cellular resolution. As a result, tissue heterogeneity and intercellular crosstalk can now be charted and delineated with never-before-seen accuracy.

Spatially annotated single-cell datasets provide unprecedented opportunities to dissect cell-cell communication. In skin biology as well as in clinical dermatology, this is particularly useful to advance our understanding of the epithelial tissue program in dermal development and disease (Cang, 2023). Epithelial organization requires several different programs that are tightly regulated. Planar cell polarity (PCP) is the organization of epithelial cells along the planar axis orthogonal to the apical-basal axis. Planar cell polarity is observed in an array of developmental processes that involve collective cell movement and tissue organization, and its disruption can lead to severe developmental defects. Recent research in flies and vertebrates has discovered new functions for planar cell polarity, as well as new signaling components and mechanistic models. However, despite this progress, the search to simplify principles of understanding continues, and important mechanistic uncertainties still pose formidable challenges. Cell migration is a highly integrated multistep process that orchestrates embryonic morphogenesis, contributes to tissue repair and regeneration, and drives disease progression in cancer, mental retardation, atherosclerosis, and arthritis. The migrating cell is highly polarized with complex regulatory pathways that spatially and temporally integrate its component processes. Neural crest cells are a well-organized, pluripotent, yet temporary group of cells from the embryonic ectoderm that have the ability to migrate and give rise to diverse cell lineages, including neurons, Schwann cells, and melanocytes (Biermann, 2022). The neuroectodermal developmental program is attributed to the elevated propensity of melanoma cells for organotropism and brain metastases.

Tissue homeostasis is supported by cellular growth and proliferation. Size is a fundamental attribute impacting cellular design, fitness, and function. Concentration-dependent cell size check points ensure that cell division is delayed until a critical target size has been achieved. Likewise, control of cell proliferation is a fundamental aspect of tissue formation in development and regeneration. Stem cells exhibit a low proliferation rate while maintaining a high proliferative capacity and are often small. In contrast, cancer cells do not precisely reverse this process yet show characteristics of loss of differentiation. Extracellular signals or ligand-receptor interactions may independently induce growth and division. Specific driver genes have been linked to specific aspects of differentiation loss and spatial organization (Hodis, 2021).

**A molecular hierarchy in melanoma**

Melanoma is an aggressive type of skin cancer that partially recapitulates the development program of normal melanocytes but in an uncontrolled and disorganized fashion. Tissue disorganization is one of the main hallmarks of cancer. Polarity proteins are responsible for the arrangement of cells within epithelial tissues through the asymmetric organization of cellular components. Consistent with these findings are recent studies investigating control elements of melanocyte and skin tissue formation (Li, 2019; Dong, 2023).

If we understand the cellular hierarchy in melanoma, wherein growth and metastasis are governed by the rules of the developing embryonic neural crest, we can delineate oncological drivers disrupting tissue homeostasis. Spatial single-cell profiling has been invaluable in quantifying the unprecedented heterogeneity and plasticity of malignant melanoma. A systems biology analysis revealed distinct subpopulations representing lineage-specific melanocyte inducing transcription factor (MITF) as well as mesenchymal-like clones with high proliferative potential (Karras, 2022; Biermann, 2022). Taken together, signaling and transcriptional elements govern a molecular hierarchy in melanoma that uncouples growth and metastasis.

### Results

**Recapitulation of a developmental program in cancer metastasis**

Early on in the history of science, the fields of evolutionary developmental biology and oncology shook hands when they created the portmanteau WNT from the planar cell polarity mutant wingless mutant drosophila and the mouse mammary tumor virus integration site. WNT-ligands are recognized by cognate frizzled (FZD) receptors. Today, the WNT, FZD, and beta-catenin signaling pathways constitute an evolutionarily conserved cell-cell communication system that is important for stem cell renewal, cell proliferation, and cell differentiation both during embryogenesis and during adult tissue homeostasis (Figure 1). In skin development, the WNT signaling pathway plays a critical role during melanocyte specification from the neural crest.

**Disruption of heterotypic signaling in skin cutaneous melanoma**

Assisted by structure-based modeling and comparative sequence alignments, cancer system biology can break down the highly conserved WNT molecular network, which encompasses 19 closely related WNT ligands (WNT1–16 including A and B isoforms) and 10 Frizzled receptors (FZD1–10) that direct the self-renewal and regeneration of many tissues during their development. FZDs are G protein-coupled receptors characterized by seven transmembrane-spanning domains, a cysteine-rich N-terminal ligand binding domain, and a C-terminal intracellular activation domain. Depending on their activation, ligands, and intracellular binding adapter proteins, FZDs are capable of transmitting extracellular signals into diverse transcriptional program outputs that determine cell fate during normal and pathogenic development.

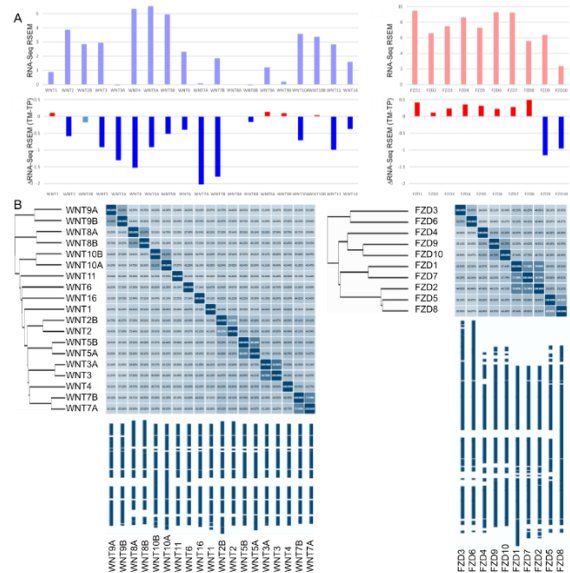

**Figure 1:** WNT-ligand (WNT) and cognate frizzled (FZD) receptor gene expression network in skin cutaneous melanoma

### Methods

RNA-Seq gene expression data of 302 patient specimens including normal, tumor,

metastatic tissue was subjected to differential gene expression analysis. Read counts were scaled via the median of the geometric means of fragment counts across all libraries. Transcript abundance was quantified using normalized single-end RNA-Seq reads in read counts as well as reads per kilobase million (RPKM). Since single-end reads were acquired in the sequencing protocol, quantification of reads or fragments yielded similar results. Statistical testing for differential expression was based on read counts and performed using EdgeR in the Bioconductor toolbox. Gene families were further analyzed using. Clustal Omega, a multiple sequence alignment program that uses seeded guide trees and Hidden Markow Model profile-profile techniques to generate alignments between sequence families. The study was carried out as part of IRB approved study dbGap ID 5094 "Somatic mutations in melanoma". The results shown are in whole based upon data generated by the TCGA Research Network http://cancergenome.nih.gov. Restricted access whole-genome sequences and whole-exome sequences were obtained from the TCGA data portal.

**Discussion**

**Planar cell polarity regulators with distinct roles in cancer metastasis**

*FZD3* and *FZD6* share high sequence homology and function through the noncanonical WNT pathway to play a critical role in migration and planar pattern formation. Structure-based sequence analysis identifies a distinct 100 amino acid C-terminal extension, exclusively found in FZD3 and FZD6. This extra tail makes FZD3 and FZD6 proteins special among their peers, enabling unique cell-cell signaling properties. Cancer genomics leverages a genome-wide insight and reveals that dysregulation of the WNT-FZD oncoprotein network is a driver of distinct spatial, developmental, and pathological programs (Figure 2).

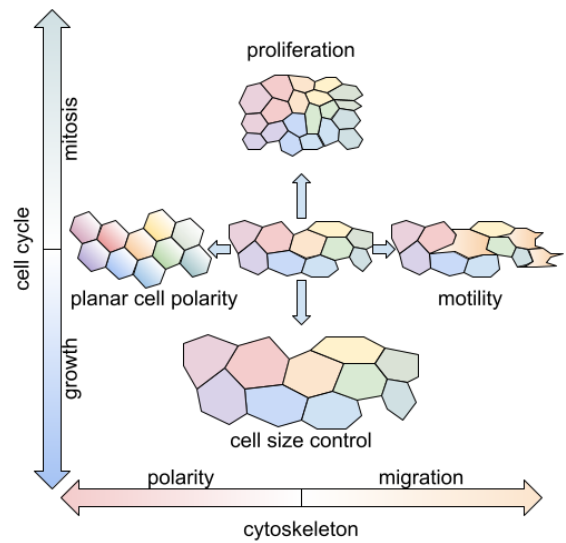

**Figure 2:** Spatial single-cell profiling identifies disruption of cellular hierarchy in cancer

*FZD3* knockout revealed the underlying transcriptional regulators of the *TWIST* and *SNAIL* families. *TWIST* and *SNAIL* are commonly defined not only as developmental markers of neural crest cells in dorsolateral migration but also as drivers of proliferation and invasion during malignant transformation. While dysregulation of *SNAIL* transcription factors is common to the *FZD6* knockout model, downregulation of epithelial mesenchymal transition (EMT)-related *ZEB* homeobox transcription factors is specific to the *FZD6* knockout model. Knockout of *FZD6* did not affect primary tumor formation or cellular proliferation but inhibited distant metastasis and cellular migration of melanoma into distant tissues. Because of similar pivotal roles in tissue polarity, sequence homology, and overexpression in melanoma, *FZD3* and *FZD6* display some functional redundancy. However, despite such similarities, the oncogenes *FZD3* and *FZD6* play different roles in melanoma progression. *FZD3* promotes melanoma metastasis by stimulating cell cycle progression, while *FZD6* regulates cell invasiveness.

In cancer, the FZD3 receptor oncoprotein signals independently of the canonical WNT pathway, while FZD6 may also engage canonical WNT signaling. The data identifies significant functional specificity of spatial regulators in cancer, despite similarity in cell planarity and tissue development. With the advent of omics approaches to spatial biology, it is now possible to link developmental programs with oncological events and explain key spatial features of metastatic aggressiveness.

## Conclusion

### Spatial omics to overcome the developmental immune privilege in disease

Facilitated by spatial omics insight, the developmental path might also have important implications for stem cell biology and oncoimmunology. Tissue aging, cellular senescence, and controlled cell death have an underappreciated immunological control element. Similarly, evasion of cancer cells from the endogenous immune repertoire can be attributed to an immune privilege that protected stem cell populations enjoy such as migrating precursors but foreign cells abuse as a hard-to-overcome escape pathway. T lymphocytes are the primary effectors of the anti-tumor response, but the interplay between melanoma and the immune system is complex, dynamic, and incompletely understood. Sustained progress in unraveling the pathogenesis of melanoma regression has led to the identification of therapeutic targets, culminating in the development of immune checkpoint inhibitors for the management of advanced disease. The first anti-FZD antibodies or engineered FZD-Fc fusion proteins, which are serving as decoy receptors for WNT ligands, are in clinical trials for patients with advanced solid tumors. Modern techniques allow for high-resolution spatial analyses of the tumor microenvironment (Filipp, 2019; Reynolds, 2021; Schäbitz, 2022). Such spatial omics studies may lead to a better understanding of the immune drivers of melanoma regression. As a result, they aid in the search for new prognostic and predictive biomarkers in the treatment of migratory metastatic cell populations (Hodis, 2022; Cang, 2023). Going forward, spatial omics will guide clinical decision-making to create durable anti-cancer therapy responses.

## Funding information

F.V.F. is grateful for the support provided by grants CA154887 from the National Institutes of Health, National Cancer Institute, Machine learning and multi-omics metabolic health around the clock, Bavaria California Technology Center (BaCaTeC), Friedrich-Alexander-University of Erlangen-Nuremberg, Erlangen, Germany, and the Science Alliance on Precision Medicine and Cancer Prevention by the German Federal Foreign Office, implemented by the Goethe-Institute, Washington, DC, USA, and supported by the Federation of German Industries (BDI), Berlin, Germany. This work is inspired by the curiosity and creativity of Franziska Violet Filipp and Leland Volker Filipp.

## Declarations

### Availability of preprint publication

The manuscript was made publically available to the scientific community on the preprint server arXiv at https://arxiv.org/abs/ and bioRxiv https://biorxiv.org/.

**Competing Interests**

There are no conflicts of interest.

**References**

bibliographyhttps://pubmed.ncbi.nlm.nih.gov/36690742/

Cang Z, Zhao Y, Almet AA, Stabell A, Ramos R, Plikus MV, Atwood SX, Nie Q. Screening cell-cell communication in spatial transcriptomics via collective optimal transport. Nat Methods. 2023 Jan 23. doi: 10.1038/s41592-022-01728-4. PMID: 36690742

https://pubmed.ncbi.nlm.nih.gov/35803246/

Biermann J, Melms JC, Amin AD, Wang Y, Caprio LA, Karz A, Tagore S, Barrera I, Ibarra-Arellano MA, Andreatta M, Fullerton BT, Gretarsson KH, Sahu V, Mangipudy VS, Nguyen TTT, Nair A, Rogava M, Ho P, Koch PD, Banu M, Humala N, Mahajan A, Walsh ZH, Shah SB, Vaccaro DH, Caldwell B, Mu M, Wünnemann F, Chazotte M, Berhe S, Luoma AM, Driver J, Ingham M, Khan SA, Rapisuwon S, Slingluff CL Jr, Eigentler T, Röcken M, Carvajal R, Atkins MB, Davies MA, Agustinus A, Bakhoum SF, Azizi E, Siegelin M, Lu C, Carmona SJ, Hibshoosh H, Ribas A, Canoll P, Bruce JN, Bi WL, Agrawal P, Schapiro D, Hernando E, Macosko EZ, Chen F, Schwartz GK, Izar B. Dissecting the treatment-naive ecosystem of human melanoma brain metastasis. Cell. 2022 Jul 7;185(14):2591-2608.e30. doi: 10.1016/j.cell.2022.06.007.

https://pubmed.ncbi.nlm.nih.gov/35482859/

Hodis E, Torlai Triglia E, Kwon JYH, Biancalani T, Zakka LR, Parkar S, Hütter JC, Buffoni L, Delorey TM, Phillips D, Dionne D, Nguyen LT, Schapiro D, Maliga Z, Jacobson CA, Hendel A, Rozenblatt-Rosen O, Mihm MC Jr, Garraway LA, Regev A. Stepwise-edited, human melanoma models reveal mutations' effect on tumor and microenvironment. Science. 2022 Apr 29;376(6592):eabi8175. doi: 10.1126/science.abi8175. Epub 2022 Apr 29. PMID: 35482859

https://pubmed.ncbi.nlm.nih.gov/36131018/

Karras P, Bordeu I, Pozniak J, Nowosad A, Pazzi C, Van Raemdonck N, Landeloos E, Van Herck Y, Pedri D, Bervoets G, Makhzami S, Khoo JH, Pavie B, Lamote J, Marin-Bejar O, Dewaele M, Liang H, Zhang X, Hua Y, Wouters J, Browaeys R, Bergers G, Saeys Y, Bosisio F, van den Oord J, Lambrechts D, Rustgi AK, Bechter O, Blanpain C, Simons BD, Rambow F, Marine JC. A cellular hierarchy in melanoma uncouples growth and metastasis. Nature. 2022 Oct;610(7930):190-198. doi: 10.1038/s41586-022-05242-7. PMID: 36131018

https://pubmed.ncbi.nlm.nih.gov/30792348/

Li C, Nguyen V, Clark KN, Zahed T, Sharkas S, Filipp FV, et al. Down-regulation of FZD3 receptor suppresses growth and metastasis of human melanoma independently of canonical WNT signaling. Proc Natl Acad Sci USA (2019) 116(10):4548–57. doi: 10.1073/pnas.1813802116658 PMID: 30792348

https://pubmed.ncbi.nlm.nih.gov/36368445/

Dong B, Simonson L, Vold S, Oldham E, Barten L, Ahmad N, et al. Frizzled 6 promotes

melanoma cell invasion but not proliferation by regulating canonical Wnt signaling and EMT. J Invest Dermatol (2022) S0022-202X(22):02660-4. doi: 10.1016/j.jid.2022.09.658 PMID: 36368445

https://pubmed.ncbi.nlm.nih.gov/25600636/

Guan J, Gupta R, Filipp FV. Cancer systems biology of TCGA SKCM: efficient detection of genomic drivers in melanoma. Sci Rep. 2015 Jan 20;5:7857. doi: 10.1038/srep07857.

https://pubmed.ncbi.nlm.nih.gov/31871830/

Filipp FV. Opportunities for Artificial Intelligence in Advancing Precision Medicine. Curr Genet Med Rep. 2019 Dec;7(4):208-213. doi: 10.1007/s40142-019-00177-4. Epub 2019 Dec 1.

https://pubmed.ncbi.nlm.nih.gov/33479125/

Reynolds G, Vegh P, Fletcher J, Poyner EFM, Stephenson E, Goh I, Botting RA, Huang N, Olabi B, Dubois A, Dixon D, Green K, Maunder D, Engelbert J, Efremova M, Polański K, Jardine L, Jones C, Ness T, Horsfall D, McGrath J, Carey C, Popescu DM, Webb S, Wang XN, Sayer B, Park JE, Negri VA, Belokhvostova D, Lynch MD, McDonald D, Filby A, Hagai T, Meyer KB, Husain A, Coxhead J, Vento-Tormo R, Behjati S, Lisgo S, Villani AC, Bacardit J, Jones PH, O'Toole EA, Ogg GS, Rajan N, Reynolds NJ, Teichmann SA, Watt FM, Haniffa M. Developmental cell programs are co-opted in inflammatory skin disease. Science. 2021 Jan 22;371(6527):eaba6500. doi: 10.1126/science.aba6500. PMID: 33479125

https://pubmed.ncbi.nlm.nih.gov/36513651/

Schäbitz A, Hillig C, Mubarak M, Jargosch M, Farnoud A, Scala E, Kurzen N, Pilz AC, Bhalla N, Thomas J, Stahle M, Biedermann T, Schmidt-Weber CB, Theis F, Garzorz-Stark N, Eyerich K, Menden MP, Eyerich S. Spatial transcriptomics landscape of lesions from non-communicable inflammatory skin diseases. Nat Commun. 2022 Dec 13;13(1):7729. doi: 10.1038/s41467-022-35319-w. PMID: 36513651